\documentclass[aps,prl,superscriptaddress,twocolumn,10pt]{revtex4-1}
\usepackage{graphicx,float}
\usepackage{bm}
\usepackage[colorlinks=true]{hyperref}
\usepackage[dvipsnames]{xcolor}
\usepackage{amsmath,amssymb}

\newcommand{\sectionn}[1]{{\textit{#1}}}
\newcommand{\reffig}[2]{\ref{#1}\,\textcolor{black}{(#2)}}
\newcommand*{\vk}{\mathbf{k}}
\newcommand*{\vn}{\mathbf{n}}
\newcommand*{\vm}{\mathbf{m}}

\begin{document}

\title{\mbox{SU(3) Topology of Magnon-Phonon Hybridization in 2D Antiferromagnets}}
\author{Shu Zhang}
\email{suzy@physics.ucla.edu}
\affiliation{Department of Physics and Astronomy, Johns Hopkins University, Baltimore, Maryland 21218, USA}
\affiliation{Department of Physics and Astronomy, University of California, Los Angeles, California 90095, USA}
\author{Gyungchoon Go}
\affiliation{Department of Materials Science and Engineering, Korea University, Seoul 02841, Korea}
\author{Kyung-Jin Lee}
\affiliation{Department of Materials Science and Engineering, Korea University, Seoul 02841, Korea}
\affiliation{KU-KIST Graduate School of Converging Science and Technology, Korea University, Seoul 02841, Korea}
\author{Se Kwon Kim}
\email{kimsek@missouri.edu}
\affiliation{Department of Physics and Astronomy, University of Missouri, Columbia, Missouri 65211, USA}
\date{April 9, 2020}

\begin{abstract}
Magnon-phonon hybrid excitations are studied theoretically in a two-dimensional antiferromagnet with an easy axis normal to the plane. We show that two magnon bands and one phonon band are intertwined by the magnetoelastic coupling through a nontrivial SU(3) topology, which can be intuitively perceived by identifying a skyrmion structure in the momentum space.
Our results are insensitive to lattice details and generally applicable to two-dimensional antiferromagnets. We show this by developing a continuum theory as the long-wavelength approximation to the tight-binding model.
The theoretical results can be probed by measuring the thermal Hall conductance as a function of the temperature and the magnetic field. 
We envision that the magnetoelastic coupling
in antiferromagnets can be a promising venue in search of various topological excitations, which cannot be found in magnetic or elastic models alone.
\end{abstract}

\maketitle

Antiferromagnets have recently emerged as promising material platforms in spintronics~\cite{JungwirthNN2016, BaltzRMP2018,Hoffmann2018Rev}. Due to the absence of the stray field and the intrinsic timescale at THz, antiferromagnet-based devices can be packed denser and operate at higher speeds than conventional ferromagnet-based GHz devices, which interact with each other via a stray field. In particular, antiferromagnets are expected to provide efficient spin-transport channels between spintronics devices, as observed
in the antiferromagnetic insulator chromia~\cite{YuanSA2018}
and haematite~\cite{LebrunNature2018}.

Parallel to the emergence of antiferromagnetic spintronics, there has been significant effort in identifying novel topological excitations in magnets, driven by both fundamental interest and their practical use as robust transport channels. Topological magnons have been identified in several types of magnets, see Refs.~\cite{ZhangPRB2013, ShindouPRB2013, ShindouPRB2013-2, MookPRB2014, KimPRL2016, OwerreJPCM2016, ZyuzinPRL2016, NakataPRB2017-2, GobelPRB2017, KimPRL2019, DiazPRL2019, KimPark2019}
for examples. Going beyond previous work on topological magnons, hybrid excitations of magnons and phonons are also shown to be topological in certain magnets, in presence of the Dzyaloshinskii-Moriya interaction (DMI)~\cite{ZhangDi2019}, the dipolar coupling~\cite{TakahashiPRL2016}, or the exchange magnetorestriction~\cite{ParkYang2019}. 
Some of us have recently shown~\cite{GoLee2019} that
a topologically nontrivial magnon-phonon hybridization can be induced in a square-lattice ferromagnet---even in the absence of inversion-symmetry-breaking DMI or long-range dipolar coupling---by a magnetoelastic coupling, which is expected to generally exist in magnets with crystalline anisotropy~\cite{Kittel1949}. 

\begin{figure}[!ht]
    \centering
    \includegraphics[width =\linewidth]{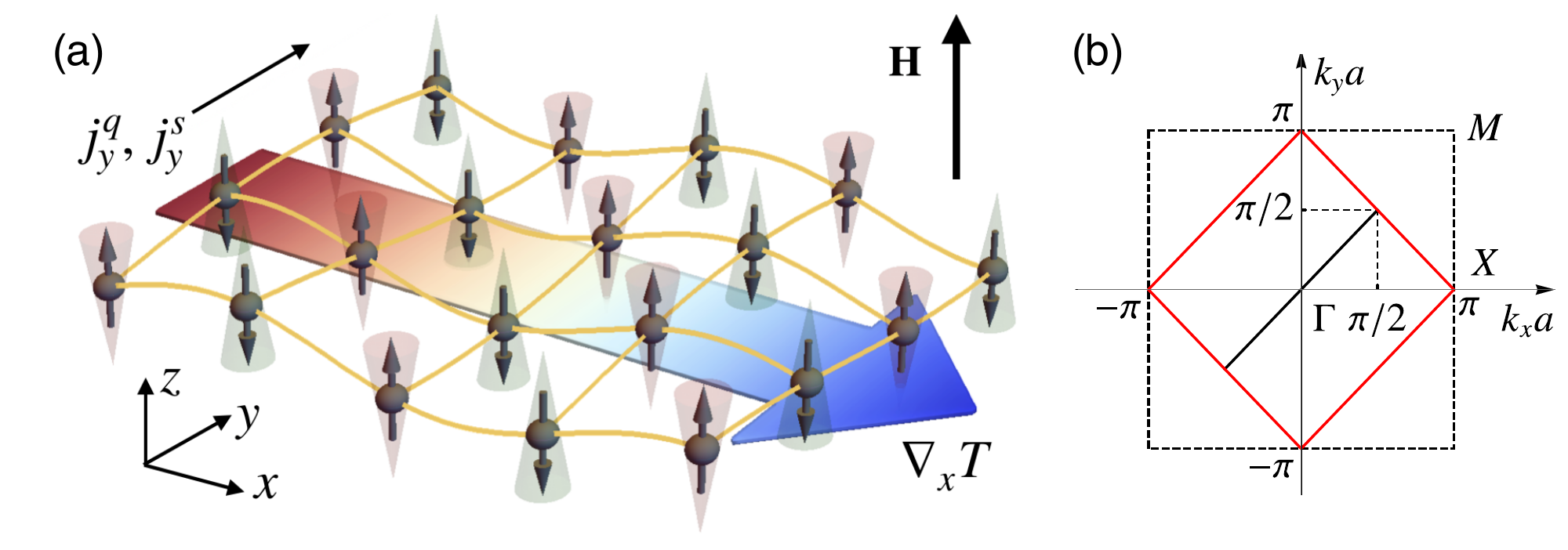}
    \caption{(a) Magnon-phonon hybrid excitations in a square-lattice antiferromagnet. The system exhibits the thermal Hall effect and the spin Nernst effect, which refer to the generation of a heat current $j^q_y$ (under magnetic field $\mathbf{H}$) and a spin current $j^s_y$ transverse to the applied temperature gradient $\nabla_x T$.
    (b) The lattice Brillouin zone (dashed black square) and the magnetic Brillouin zone (red square), where $a$ is the lattice constant.}
    \label{fig:lattice}
    \vspace{-10pt}
\end{figure}

In this Letter, we address the topological effect of the magnetoelastic coupling~\cite{Kittel1958} in a square-lattice \textit{antiferromagnet}, where
a minimal model to describe 
magnon-phonon hybrid excitations 
requires three quasiparticles. Different from the SU(2) topology of the ferromagnetic case with two bands~\cite{GoLee2019}, they possess an enriched topological structure of the SU(3) algebra.
Aside from directly calculating Chern numbers of excitations bands using the SU(3) formalism,
we show that, an intuitive understanding of the band topology can be obtained by identifying a skyrmion structure, where skyrmion numbers are defined through the more familiar SU(2) formalism~\cite{qi2011review}.
The nontrivial topology gives rise to the thermal Hall effect, where the direction of the transverse heat flow can be controlled, as the thermal Hall conductance changes sign upon reversal of the out-of-plane magnetic field, see Fig.~\ref{fig:lattice}~(a). There has been growing interest in a family of two-dimensional antiferromagnets $M$PS$_3$ ($M$ = Mn, Fe, Ni)~\cite{Burch2018review,gibertini2019}, where we propose the thermal Hall effect to be observed due to the magnetoelastic coupling discussed in this work. For experimental comparison, we give a prediction for the dependence of the thermal Hall conductance on the temperature and the magnetic field.
Furthermore, by developing a continuum field theory for the magnetic and elastic degrees of freedom, we show that our results are generally applicable to two-dimensional antiferromagnets.
We envision antiferromagnets serving as versatile platforms to realize various multiband topological insulators, where magnon-phonon excitations can be used for robust information transport in spintronics. The broader implications of our work are discussed at the end of the Letter.

\sectionn{Model}|We study harmonic excitations of the Hamiltonian 
\begin{equation}
    H = H_\textit{m} + H_\textit{e} + H_\textit{me}
\label{eq:full-Hamiltonian}
\end{equation} 
on a square lattice, which describes two subsystems---magnetic and elastic---and the interaction between them.

The magnetic interactions include the nearest-neighbor antiferromagnetic Heisenberg exchange, the easy-axis anisotropy, and the Zeeman coupling to an external magnetic field $\mathbf{H} \!=\! H \hat{\mathbf{z}}$:
\begin{equation}
    H_\textit{m} 
    = J \sum_{ \langle \ell \ell' \rangle} \mathbf{S}_\ell \cdot \mathbf{S}_{\ell'}
    - \frac{K}{2} \sum_\ell S_\ell^{z2}
    - h \sum_\ell S_\ell^z,
\label{eq:magnetic-Hamiltonian}
\end{equation}
where 
$\langle \ell \ell' \rangle$ runs over all pairs of nearest neighbors, $J, K>0$ are respectively the exchange constant and the anisotropy coefficient,
and the Zeeman coupling is $h \!=\! \mu_B g H$ with the Bohr magneton $\mu_B$
and the Land{\'e} g factor, $g$.
One of the two time-reversal-related N{\'e}el ground states has the spin configuration $\mathbf{S}_{i \in A} \!=\! S \hat{\mathbf{z}}$ and $\mathbf{S}_{j \in B} \!=\! -S \hat{\mathbf{z}}$, as shown in Fig.~\reffig{fig:lattice}{a}.
We study magnon excitations by the Holstein-Primakoff approach~\cite{HolsteinPR1940},
\begin{equation}
\begin{aligned}
     H_\textit{m} = 
     \sum_{\vk} \left( \epsilon_\vk^{m+} \alpha_\vk^\dagger \alpha_\vk 
     + \epsilon_\vk^{m-} \beta_\vk^\dagger \beta_\vk \right) + \text{const.},
\end{aligned}
\end{equation}
where the wave vector $\vk$ is summed over the first magnetic Brillouin zone shown in Fig.~\reffig{fig:lattice}{b}. 
There are two magnon bands with dispersion relations 
$ \epsilon_\vk^{m\pm}  \!=\! S \sqrt{(z J + K)^2 \!-\! (J \gamma_\vk)^2} \!\pm\! h $,
where $z=4$ is the coordinate number and $\gamma_\vk \!=\! 2 \left[ \cos ( k_x a) \!+\! \cos (k_y a) \right]$.
The bosonic operator $\alpha_\vk^\dagger$ ($\beta_\vk^\dagger$) creates a magnon with $S^z = -1$ ($+1$), which is a good quantum number because of the rotational symmetry of the Hamiltonian~(\ref{eq:magnetic-Hamiltonian}). We assume $h\lesssim S\sqrt{8JK}$ so that magnons have finite gaps.

For the Hamiltonian describing the lattice dynamics, we consider the out-of-plane component $u_\ell^z$ of the displacement vector $\mathbf{u}_\ell$, which is the only component that later enters the magnon-phonon coupling:
\begin{equation}
    H_\textit{e} = \frac{1}{2M} \sum_\ell (p^z_\ell)^2 
    + \frac{\lambda}{2} \sum_{ \langle \ell \ell' \rangle}  (u^z_\ell - u^z_{\ell'})^2,
\label{eq:phonon-Hamiltonian}
\end{equation}
where $M$ is the atom mass,  $p^z_\ell \!=\! M \dot{u}^z_\ell$ is the conjugate momentum, and $\lambda$ is the spring constant between nearest neighbors.
The quantization of phonon excitations yields
\begin{equation}
    H_\textit{e} = \sum_{\vk} \left( \epsilon_\vk^{p-} \eta_\vk^\dagger \eta_\vk 
    +  \epsilon_\vk^{p+} \zeta_\vk^\dagger \zeta_\vk \right) + \text{const.},
\end{equation}
where phonon dispersion relations are $\epsilon_\vk^{p\pm} \!=\! \hbar \omega_0 \sqrt{z \pm \gamma_\vk} $
with characteristic frequency $\omega_0 \!=\! \sqrt{\lambda /M}$.
In the reduced (magnetic) Brillouin zone, we obtain two phonon modes, the acoustic branch $\eta_\vk$ and the optical branch $\zeta_\vk$.

\begin{figure}[!ht]
    \centering
    \includegraphics[width = \linewidth]{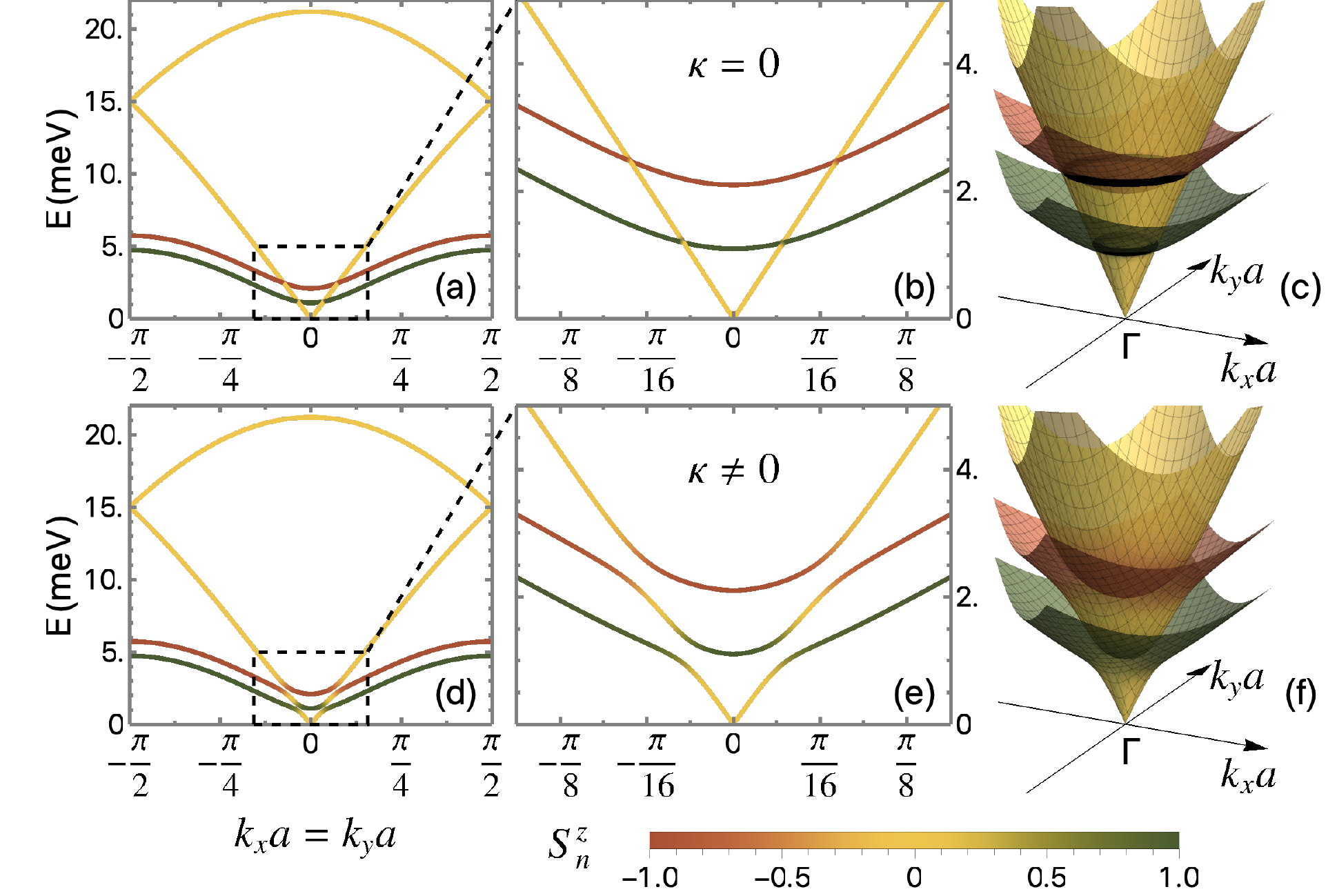}
    \caption{Dispersion relations of excitation bands. (a)--(c) In the absence of the magnetoelastic coupling. The green and the red lines depict the $S^z \!=\! 1$ magnons and the $S^z \!=\! -1$ magnons, respectively. The yellow lines depict phonons. (d)--(f) Hybridization of magnons and the acoustic phonon in the presence of the magnetoelastic coupling $\kappa = 0.8$~meV/{\AA}. The color represents the value of $S^z_n$ of the $n$th band at each momentum. Parameters used are these: 
    $S = 5/2$,
    $J \!=\! h \!=\! 0.5$~meV, 
    $K \!=\! 0.1$~meV, 
    and $\hbar \omega_0 \!=\! 7.5$~meV.
    }
\vspace{-5pt}
\label{fig:bands}
\end{figure}

\begin{figure*}[!ht]
    \centering
    \includegraphics[width = 0.8\linewidth]{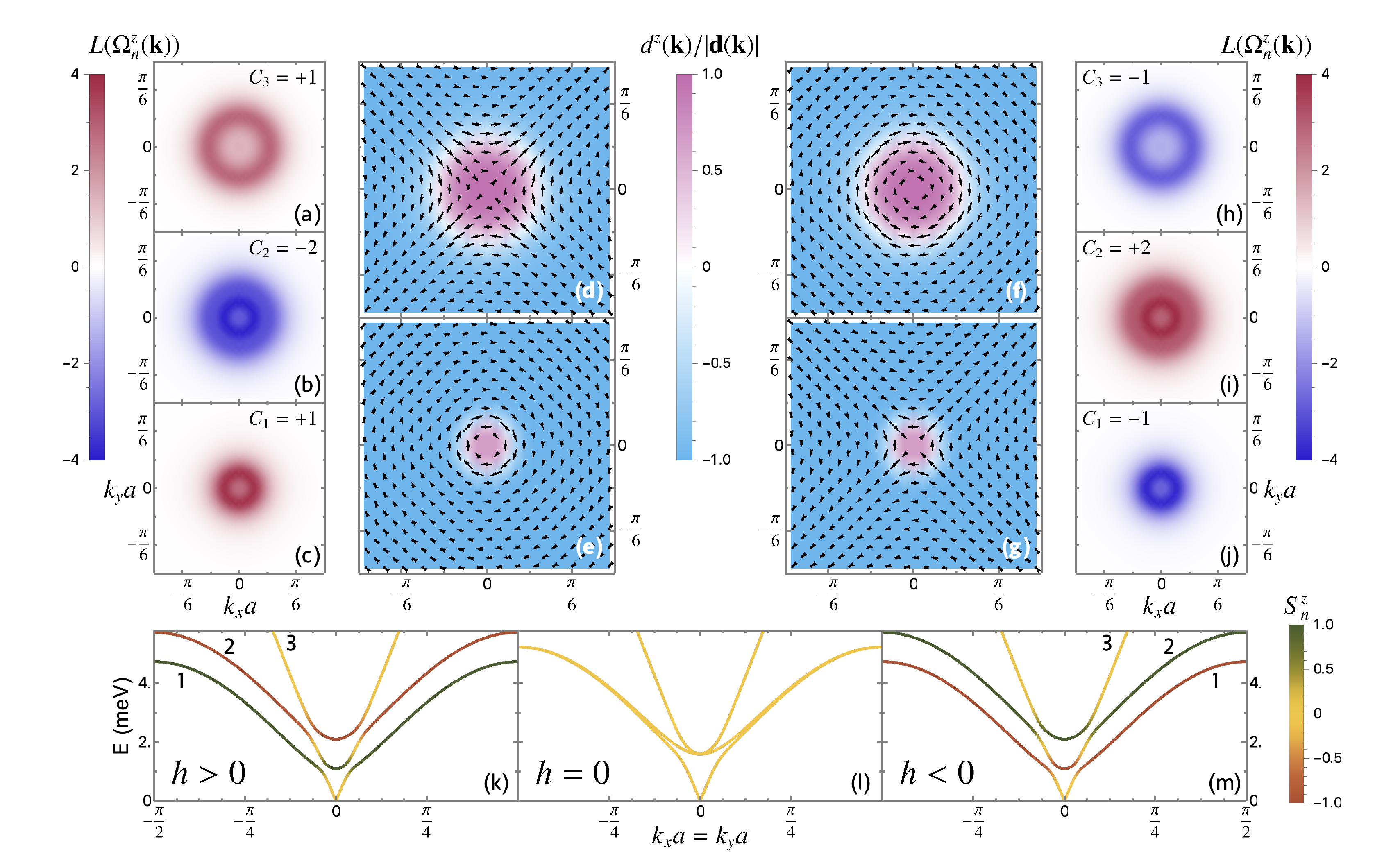}
    \caption{Topological properties of magnon-phonon hybrid excitations. Berry curvatures $\Omega_n^z$ of excitation bands for $h>0$ (a)--(c) and $h<0$ (h)--(j) are plotted in log scale~\cite{GoLee2019}
    $L(\Omega_n^z) = {\rm sign}(\Omega_n^z) {\rm log} (1+|\Omega_n^z|)$.  The corresponding skyrmion structures are shown for $h>0$ (d), (e) and $h<0$ (f), (g), where black arrows are in-plane components of $\mathbf{d}(\vk)$. Same parameters are used as in Fig.~\ref{fig:bands}.
    A topological phase transition (k)--(m) occurs at $h=0$ when band gaps close. See the main text for detailed discussions.
    }
\vspace{-10pt}
\label{fig:skyrmions}
\end{figure*}

Our study focuses on the interaction between the spin and the elastic strain, which will be shown later to induce nontrivial topology to the excitation bands. As was pointed out by Kittel~\cite{Kittel1958},
\begin{equation}
    H_{\textit{me}}  = \kappa \sum_{\ell \in A, B} \sum_{\delta} S_\ell^z
    (\mathbf{S}_\ell \cdot \hat{\mathbf{e}}_\delta) (u^z_\ell - u^z_{\ell+\delta}),
\label{eq:me-Hamiltonian}
\end{equation}
where $\kappa$ is the magnetoelastic coupling constant and $\hat{\mathbf{e}}_\delta $ is a unit vector pointing along the bond connecting nearest neighbors. For our magnetic ground state, this term~(\ref{eq:me-Hamiltonian}) is the leading order approximation to the magnetoelastic energy.

When $\kappa \!=\! 0$, the acoustic phonon intersects magnon bands at rings of wave vectors as shown in Figs.~\reffig{fig:bands}{a--c}.  
We treat the magnetoelastic coupling as a perturbation, the most prominent effect of which is to lift the degeneracy at band crossings.
Therefore, the magnetoelastic Hamiltonian~(\ref{eq:me-Hamiltonian}) expressed in magnon and phonon operators is dominated by terms conserving the total number of quasiparticles,
\begin{equation}
\begin{aligned}
    H_\textit{me}  & \approx \kappa   \sum_{\vk} 
    \sum_{b = +,-} \Big\{ v_\vk^b \chi^b_\vk
    \left[i \sin (k_x a) 
    (\alpha_\vk^\dagger  +b \beta_\vk^\dagger)  \right. \\
    &- \left.  \sin (k_y a)  
    (\alpha_\vk^\dagger  -b \beta_\vk^\dagger)\right]
    + \text{H.c.}\Big\},
\label{eq:leading-me}
\end{aligned}
\end{equation}
where $\chi^+_\vk = \zeta_\vk$, $\chi^-_\vk = \eta_\vk$,
$v_\vk^\pm \!=\!   \hbar S^{3/2}  \sqrt{1/ 2M \epsilon_\vk^{p\pm}}\times \left(\sinh \vartheta_\vk \pm \cosh \vartheta_\vk  \right)$ with
$\tanh 2 \vartheta_\vk = -J \gamma_\vk/(zJ \!+\! K) $, .

The full Hamiltonian~(\ref{eq:full-Hamiltonian})
thus becomes a tight-binding model for magnon-phonon hybrid excitations. By exact diagonalization, we find that the magnetoelastic coupling gives rise to the so-called avoided crossing features in excitation bands, as shown in Figs.~\reffig{fig:bands}{d--f}. The degree of hybridization is reflected by the spin number
$S_n^z (\vk) \!=\! \langle \phi_n (\vk)| 
(\beta_\vk^\dagger \beta_\vk \!-\!\alpha_\vk^\dagger \alpha_\vk) 
|\phi_n (\vk) \rangle$, 
where $\phi_n (\vk)$ is the wave function of the band with the eigenenergy $E_n (\vk)$.

\sectionn{SU(3) Topology}|To study the band topology, it is adequate to consider the three-band Hamiltonian~\cite{supmat}
$ H_3 \!=\! \sum_\vk \psi_\vk^\dagger \mathcal{H}_\vk \psi_\vk$, with the operator
$\psi_\vk \!=\! (\alpha_\vk, \beta_\vk, \eta_\vk)^T$ and the hopping matrix
\begin{equation}
    \mathcal{H}_\vk = \left( \begin{array}{ccc}
    \epsilon_\vk^{m+} & 0 & C_\vk  \\
    0 & \epsilon_\vk^{m-} & C_\vk^* \\
    C_\vk^* & C_\vk & \epsilon_\vk^{p-} \\
    \end{array} \right) ,
\label{eq:three-band-Hk}
\end{equation}
with the magnon-phonon coupling $C_\vk \!=\! \kappa v_\vk^- \left[i \sin (k_x a) \! \right.$ 
$-  \left. \! \sin (k_y a) \right]$.
The three-band tight-binding model is naturally described by the algebra of the SU(3) group. By making a similarity transformation and eliminating the trace, the Bloch Hamiltonian~(\ref{eq:three-band-Hk}) becomes
\begin{equation}
\begin{aligned}
    \widetilde{\mathcal{H}}_\vk 
    &=  h \lambda_2 
    + \sqrt{2} \kappa   v_\vk^-  \left[ \sin (k_x a) \lambda_5 - \sin (k_y a) \lambda_7 \right] \\
    & +\frac{1}{2\sqrt{3}}(\epsilon_\vk^{m+} + \epsilon_\vk^{m-} - 2\epsilon_\vk^{p-}) \lambda_8 ,
\label{eq:H-in-lambda}
\end{aligned}
\end{equation} 
where $\lambda$s are Gell-Mann matrices~\cite{Gell-MannPR1962} for SU(3) group akin to Pauli matrices for SU(2) group.

Following the method developed by~\textcite{BarnettGalitski2012}, we compute the Berry curvature $\Omega^z_n(\vk)$ and the Chern number $C_n$ associated with each band directly from the Bloch Hamiltonian~(\ref{eq:H-in-lambda}).
For $h>0$, we find the Chern numbers to be $( +1, - 2,  +1)$, see Figs.~\reffig{fig:skyrmions}{a--c}.
All Chern numbers flip their signs when the magnetic field is in the opposite direction, see Figs.~\reffig{fig:skyrmions}{h--j} for $h<0$.
As shown in Figs.~\reffig{fig:skyrmions}{k--m}, a topological phase transition happens at $h=0$, when the band gaps close at both $\Gamma$ point and the boundary of the Brillouin zone.

We notice that $\left( \lambda_2, \lambda_5, \lambda_7 \right)$ forms the spin-1 representation of angular momentum operators. Three-band hopping models enjoying such a SU(2) subalgebra have been well understood\cite{OhgushiNagaosa2000,HeVarma2012,GoHan2013}, where Chern numbers are $(-c,0,c)$ with $c \in \mathbb{Z}$. 
The presence of $\lambda_8$ enriches the band topology of our model.
Similar cases have been previously studies in cold atom systems~\cite{BarnettGalitski2012,UedaShannon}, where formalism with more mathematical rigor is discussed. 

Here, we achieve an understanding of the band topology more intuitively, with the help of the skyrmion structure in the familiar two-band theories.
For $\kappa a \! \ll \! h $, nonzero Berry curvatures concentrate in the vicinity of avoided crossings, while the two rings of momenta satisfying
$\epsilon_\vk^{m\pm} = \epsilon_\vk^{p-}$ are relatively separated from each other. 

Let us take the example of $h\!>\!0$. For wave vectors near the band crossing $\epsilon_\vk^{m-} \!=\! \epsilon_\vk^{p-}$, the prominent topological effect should occur between the two crossing bands, which motivates us to write the Bloch Hamiltonian~(\ref{eq:three-band-Hk}) in the following form:
\begin{equation}
   \mathcal{H}_\vk = \left( \begin{array}{cc}
   \epsilon_\vk^{m+}  & 0   \\
    0 & 
    \frac{1}{2}(\epsilon_\vk^{m-} + \epsilon_\vk^{p-}) I_2 
    + \mathbf{d}(\vk) \cdot \bm{\sigma} \\
    \end{array} \right) + \mathcal{V}_\vk,
\end{equation}
where $\bm{\sigma} \!=\! (\sigma^x,\sigma^y,\sigma^z)$ are Pauli matrices,
$\mathbf{d}(\vk) \!=\! \left[ -\kappa   v_\vk^- \sin (k_y a),
\kappa   v_\vk^- \sin (k_x a),
(\epsilon_\vk^{m-} \!-\! \epsilon_\vk^{p-})/2 \right]$.
Here, $\mathcal{V}_\vk$ is a perturbation term that does not participate in opening the gap between the two bands ($\epsilon_\vk^{m-}$ and $\epsilon_\vk^{p-}$) and thus can be neglected for now. A skyrmion with topological charge $Q \!=\! +1$ [see Fig.~\reffig{fig:skyrmions}{e}] emerges in the unit vector field 
$\hat{\mathbf{d}}(\vk) \! = \!\mathbf{d}(\vk)/|\mathbf{d} (\vk)|$, where 
\begin{equation}
    Q = \frac{1}{4\pi} \int \! d k_x \! \int \! d k_y \;
    \hat{\mathbf{d}}(\vk) \cdot
    \left( \frac{\partial  \hat{\mathbf{d}}(\vk)}{\partial k_x}
    \times 
    \frac{\partial  \hat{\mathbf{d}}(\vk)}{\partial k_y}
    \right).
\end{equation}
According to the SU(2) model for two bands gapped by $\mathbf{d}(\vk)$, the lower (upper) band has Chern number $+Q$ ($-Q$). 
Similarly, around 
$\epsilon_\vk^{m+} \!=\! \epsilon_\vk^{p-}$, the Berry curvature is determined by an antiskyrmion with $Q \!=\! -1$ [see Fig.~\reffig{fig:skyrmions}{d}]. Together, the skyrmion and anti-skyrmion structure gives a Chern number distribution $(+1, -2, +1)$. 
For $h<0$, the only difference lies at $\epsilon_\vk^{m-} \!>\! \epsilon_\vk^{m+}$. The Chern numbers are thus reversed, with the skyrmion structure shown in Figs.~\reffig{fig:skyrmions}{f,~g}.
The consideration of the skyrmion structure is consistent with the homotopy theory for an SU(3) matrix with a fixed set of eigenvalues~\cite{Nakahara2003,LeeHan2015su3}:
$\pi_2 ({\rm SU}(3)/({\rm U}(1) \times {\rm U}(1)))
\simeq \pi_1 ({\rm U}(1)) \times \pi_1 ({\rm U}(1)) = \mathbb{Z} \times \mathbb{Z}$.




\begin{figure}[!b]
    \centering
    \includegraphics[width = \linewidth]{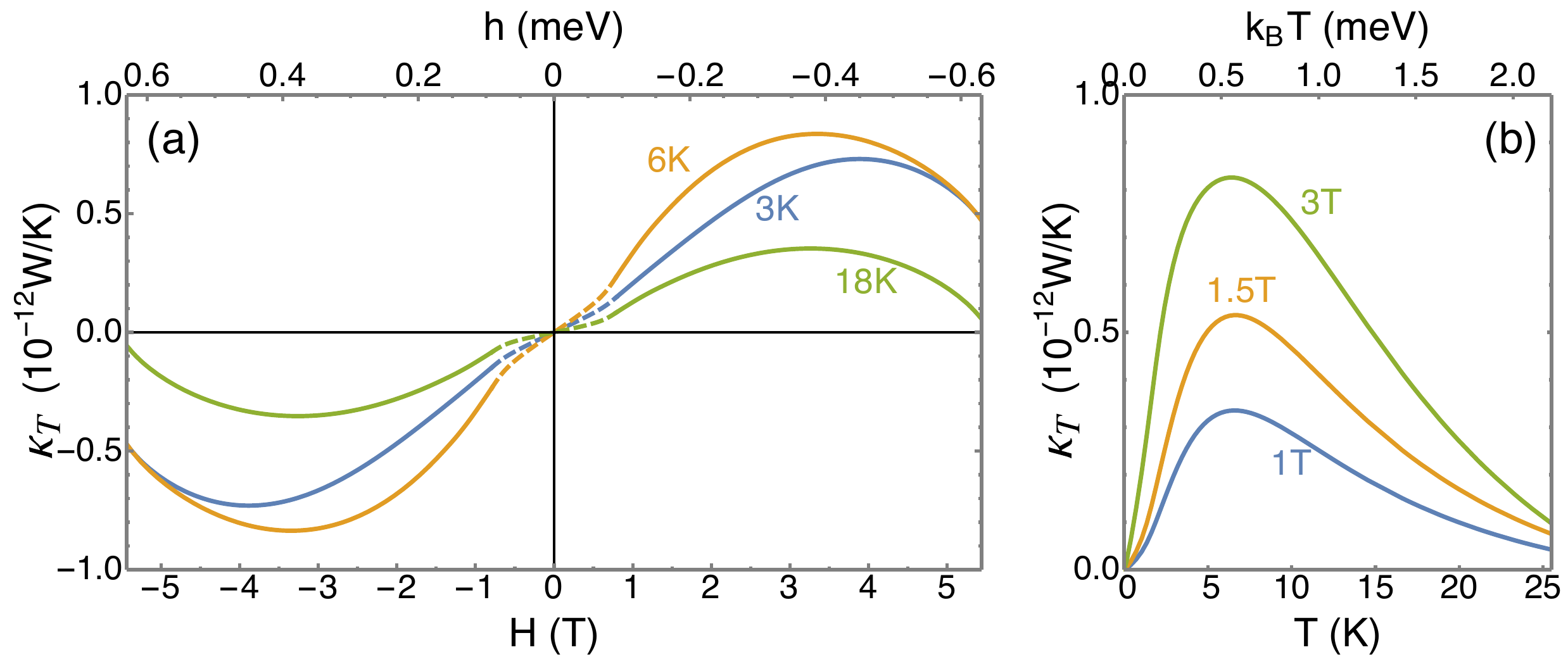}
    \caption{The magnetic field dependence (a) and temperature dependence (b) of the thermal Hall conductance $\kappa_T$. The dashed lines are extrapolations from calculation. Parameters and estimated for MnPS$_3$, see main text.}
\vspace{-10pt}
\label{fig:Hall}
\end{figure}

\sectionn{Experimental applications}|The nontrivial band topology gives rise to thermal Hall effect, which can be observed in experiments. 
A current of magnon-phonon excitations drifting under a temperature gradient acquires a transverse component in presence of the fictitious magnetic field $\Omega_n^z(\vk)$. The thermal Hall conductance 
is given by~\cite{Matsumoto2011PRL,matsumoto2011}
$\kappa_T \!=\! -(k_B^2 T/\hbar)
\sum_{n} \int d^2 \vk \,  \Omega^z_n(\vk) c_2 [\rho_n(\vk)]$,
where $c_2 (\rho) =  (1+\rho) \ln^2 [ (1+\rho)/\rho]- \ln^2 \rho - 2 {\rm Li}_2(-\rho)$, $\rho_n(\vk) = 1/[\exp(E_n(\vk)/{k_B T}) - 1]$ is the Bose-Einstein distribution function, and ${\rm Li}_2 (z)$ is the polylogarithm function.

We estimated a set of parameters for MnPS$_3$ to plot the temperature $T$ and magnetic field $H$ dependence of the thermal Hall conductance in Fig.~\ref{fig:Hall}. A Mn ion has mass $M \!=\! 55$~u and spin $S \!=\! 5/2$. The lattice constant is $a \!=\! 6.1$~{\AA}. We set $\hbar \omega_0 \!=\! 7.5$~meV, resulting in a phonon velocity $\sim 5$~km/s; magnetic parameters~\cite{Joy1992,Wildes1998} $zJ = 4$~meV and $K \!=\! 0.01$~meV, yielding a small magnon gap $\sim 0.7$~meV; the Land{\'e} g factor $g \!=\! -2.0$; and the magnetoelastic coupling $\kappa \!=\! 0.2$~meV/{\AA}, giving $2\kappa S^2/a^2 \sim 11$~J/cm$^3$, in the same order of magnitude with Kittel's estimation~\cite{Kittel1949}. 
The resulting thermal Hall conductance has a magnitude comparable to the observed values in Ref.~\cite{banerjee2018}, thus is within the current experiment reach. 

The antisymmetric feature of $\kappa_T$ under the magnetic field can be of practical interest in designing spincaloritronic devices with reversal thermal current.
Since the magnetoelastic coupling explicitly couples the spin orientation and the lattice wave vector [see Eq.~(\ref{eq:continuum-me}) in the continuum theory], the spin Nernst effect~\cite{BauerNM2012,cheng2016Nernst} is also expected to be observable. 
According to the bulk-edge correspondence~\cite{murakami2016,thingstad2019}, there are two edge modes carrying a net spin current.

\sectionn{Applicability to general 2D antiferromagnets}|The 
nontrivial SU(3) band topology in magnon-phonon hybrid excitations is not limited to a square-lattice system. We expect it to be a rather general 
phenomenon
in two-dimensional antiferromagnets. This is demonstrated by developing a continuum theory, which is equivalent to the tight-binding model in the low-energy and long-wavelength limit (with $\kappa a \ll S\sqrt{2J^3/K} \ll \hbar \omega_0$).

We study the dynamics of two slowly varying fields---the unit-vector staggered spin
$ \vn (\mathbf{r},t) \!=\! (\mathbf{S}_A - \mathbf{S}_B)/2S \!=\! \hat{\mathbf{z}} + \delta \vn (\mathbf{r},t) $~\cite{kim2014propulsion}, and the lattice displacement $u^z(\mathbf{r},t)$. They are coupled through the energy density
\begin{equation}
    \mathcal{H}_\textit{me} = -\frac{2 \kappa S^2}{a}  \vn \cdot \nabla u^z.
\label{eq:continuum-me}
\end{equation}
Dispersion relations of the magnon-phonon hybrid excitations are revealed by their coupled equations of motion, which can be derived from the action 
\begin{equation}
    \mathcal{S} [\Psi]
    \approx  \frac{1}{2} \int {d^2 \vk} \int {d \omega}  \;
    \Psi_{\vk, \omega}^\dagger 
    \left( E - \mathcal{G}_{\vk,\omega} \right)
    \Psi_{\vk, \omega},
\label{eq:action-continuum-theory}
\end{equation}
where the operator $\Psi_{\vk, \omega}$ is the Fourier transform of the field operator
$\left[ \psi^+ (\mathbf{r},t), \, \psi^- (\mathbf{r},t), \, u^z (\mathbf{r},t) \right]^T$ with 
$\psi^\pm \!=\! \delta n_x \pm i \delta n_y $; and 
\begin{equation}
    \mathcal{G}_{\vk,\omega} =  
        \left(\begin{array}{ccc}
        E_m+h & 0 & \widetilde{C}_\vk \\
        0 & E_m-h & \widetilde{C}_\vk^* \\
        \widetilde{C}_\vk^* & \widetilde{C}_\vk & E_p
        \end{array} \right)
\label{eq:propagator}
\end{equation}
with $E_m \!=\! S\sqrt{8JK} \!+\! S \sqrt{2J^3/K} a^2 (k_x^2 + k_y^2)$, 
$E_p \!=\! \hbar \omega_0 a \sqrt{k_x^2 + k_y^2}$, and $\widetilde{C} \!=\! -\kappa S^{3/2} (8J/K)^{1/4}\sqrt{a \hbar/2 M \omega_0} \times \, (k_x^2 + k_y^2)^{-1/4} (ik_x \!-\! k_y)$. These are exactly the lowest order approximations to elements in the Block Hamiltonian~(\ref{eq:three-band-Hk}), and thus the continuum theory~(\ref{eq:action-continuum-theory}) possesses the same band topology topology with that of square-lattice system: Our results on the nontrivial SU(3) topology of magnon-phonon hybrid excitations and the resultant thermal Hall conductivity are expected to exist generally in two-dimensional antiferromagents regardless of their lattice structures.


\sectionn{Discussion}|We remark on the applicability of our theoretical model more generally. First, we study a classical spin system with an antiferromagnetic order. At a temperature is much lower than the N{\'e}el temperature, magnetic excitations are magnons with well-defined dispersions. Second, we have focused on the hybridization between two magnon bands and the acoustic phonon, which is a minimal model for an antiferromagnet. The effect of the magnetoelastic coupling can, of course, be generally considered for more complicated band structures, such as phonons from other displacement components and magnons in noncollinear antiferromagnet. Third, the rotational symmetry around the easy axis is approximate in the low energy limit in a lattice system. 

Several features of the magnetoelastic coupling~(\ref{eq:me-Hamiltonian}) are worth noting in our study.
It is linear in magnon operators, as manifested in Eq.~(\ref{eq:leading-me}), thus has a leading order contribution to the excitation spectrum. 
In the continuum form~(\ref{eq:continuum-me}), it has the familiar form of a spin-orbit coupling. Without the requirement of breaking the inversion symmetry, it is expected to exist generally in magnetic systems. In contrast to the widely studied Dzyaloshinskii-Moriya interaction~\cite{DZYALOSHINSKY1958,Moriya1960}, the effect of the magnetoelastic coupling~(\ref{eq:me-Hamiltonian}) is largely unknown. By revealing the nontrivial topology it can induce to magnon-phonon hybrid excitations, we hope to bring more attention to its effects.

The magnetoelastic coupling can serve as a venue in search of various topological excitations, as well as other novel phenomena in the interplay between magnetic and elastic degrees of freedom. 
For example, nonreciprocal phonons in a magnetic ordered state (as recently observed in Ref.~\cite{NomuraSeki2019}) can be a more generally existing effect~\cite{Kittel1958}; and vice versa, the possibility to use strain to tune magnetic states or mangon properties can be considered~\cite{Burch2018review}.

An interesting subject for future study is the effect from the intrinsic quantum mechanical nature of antiferromagnetic magnons, which can be prominent when the spin length is small. Another is the magnon-phonon hybridization engendered by the exchange-type magnetoelastic coupling as in Ref.~\cite{ParkYang2019}, which could be strong especially in materials exhibit the Jahn-Teller effect~\cite{Sushkov2005}.
It has been shown that magnon-photon coupling can result similar band repelling effect~\cite{ManoharVenkataraman1972,BoseZuniga1975}. Our ideology might also be of interest for topological photonics~\cite{Ozawa2019TopoPhotonics} in magnetic systems.

\begin{acknowledgments}
We thank Oleg Tchernyshyov for comments on the manuscript. S. Z. and S. K. K. acknowledge the support by the University of Missouri. S. Z. is also supported by the U.S. DOE Basic Energy Sciences,
Materials Sciences and Engineering
Award No. DE-SC0019331. S. K. K. is supported by Young Investigator Grant (YIG) from Korean-American Scientists and Engineers Association (KSEA).
G. G. is supported by the NRF of Korea (Grant No. NRF--2019R1I1A1A01063594). K.-J. L. acknowledges a support by the National Research Foundation (NRF) of Korea (Grant No. NRF-2017R1A2B2006119).
\end{acknowledgments}


\bibliographystyle{apsrev4-1}
\bibliography{MagnonPhonon}

\onecolumngrid
\clearpage
\setcounter{equation}{0}
\renewcommand{\theequation}{S\arabic{equation}}
\appendix

{\centering
    \large{\textbf{{Supplementary Material}}}
\par}

\section{Definitions of bosonic operators}

To find magnon excitations in the magnetic Hamiltonian~(\ref{eq:magnetic-Hamiltonian}), we perform the Holstein-Primakoff transformation: 
\begin{equation}
\begin{aligned}
i \in A: & \;
S_i^+ = \sqrt{2S - n_i} a_i, \quad 
S_i^- = a_i^\dagger \sqrt{2S - n_i} , \quad 
S_i^z = S - n_i, \quad
n_i = a_i^\dagger a_i ;\\
j \in B: & \;
S_j^+ = b_j^\dagger \sqrt{2S - n_j} , \quad 
S_j^- = \sqrt{2S - n_j} b_j, \quad 
S_j^z = - S + n_j, \quad
n_j = b_j^\dagger b_j .
\end{aligned}
\end{equation}
The Fourier transform is taken as follows
\begin{equation}
\begin{aligned}
a_\vk &= \sqrt{\frac{2}{N}} \sum_{i \in \text{I}} e^{-i \vk \cdot \mathbf{r}_i} a_i; \quad 
b_\vk &= \sqrt{\frac{2}{N}} \sum_{j \in \text{II}} e^{-i \vk \cdot \mathbf{r}_j} b_j,
\end{aligned}
\end{equation}
where $N$ is the total number of lattice sites. The Bogoliubov transformation is given by
\begin{equation}
a_\vk \!=\! \alpha_\vk \cosh \vartheta_\vk  \!+\! \beta_{-\vk}^\dagger \sinh \vartheta_\vk, \quad
b_\vk \!=\! \beta_\vk \cosh \vartheta_\vk \!+\! \alpha_{-\vk}^\dagger \sinh \vartheta_\vk,
\end{equation}
with 
$\tanh 2 \vartheta_\vk \!=\! -J \gamma_\vk/(zJ \!+\! K)$.

The phonon operators are 
\begin{equation}
\begin{aligned}
\eta_\vk &=  \frac{1}{\sqrt{N}}\sum_{\ell}  \sqrt{\frac{M \omega_\vk^{p-}}{2 \hbar}}
\left( \frac{i}{ M \omega_\vk^{p-}} \, p^z_\ell e^{i \vk \cdot \mathbf{r}_\ell}  
+  u^z_\ell e^{-i \vk \cdot \mathbf{r}_\ell} \right), \\
\zeta_\vk &=  \frac{1}{\sqrt{N}} \sum_{\ell}  \sigma_\ell \sqrt{\frac{M \omega_\vk^{p+}}{2 \hbar}}
\left( \frac{i}{ M \omega_\vk^{p+}} \, p^z_\ell e^{i \vk \cdot \mathbf{r}_\ell}  
+   u^z_\ell e^{-i \vk \cdot \mathbf{r}_\ell} \right),
\end{aligned}
\end{equation}
where $\omega_\vk^{p\pm} = \omega_0 \sqrt{2\pm \gamma_\vk}$, $\sigma_{i \in A} = 1$ and $\sigma_{j \in B} = -1$.

\section{Justification to neglect the optical phonon}

In the main text, we have studied the three-band Bloch Hamiltonian~(\ref{eq:three-band-Hk}), neglecting the optical phonon, which is energetically separated from the other bands for momenta far away from the boundary of the Brillouin zone. We now show that this does not affect the discussion on the band topology.

The full Bloch Hamiltonian is 
\begin{equation}
\mathcal{H}'_\vk = \left( \begin{array}{cccc}
\epsilon_\vk^{m+} & 0 & M_1 & M_2^* \\
0 & \epsilon_\vk^{m-} & M_1^* &  - M_2 \\
M_1^* & M_1 & \epsilon_\vk^{p-} & 0 \\
M_2 & - M_2^* & 0 & \epsilon_\vk^{p+}
\end{array} \right)
\label{eq:four-band-Hk}
\end{equation}
where 
\begin{equation}
\begin{aligned}
M_1 &= C_\vk = -\kappa \hbar S^{3/2} \sqrt{\frac{1}{2M \epsilon_\vk^{p\pm}}}  \left( \cosh \vartheta_\vk  - \sinh \vartheta_\vk \right) \left[i \sin (k_x a) - \sin (k_y a)\right], \\
M_2 &= - \kappa \hbar S^{3/2} \sqrt{\frac{1}{2M \epsilon_\vk^{p\pm}}} \left( \cosh \vartheta_\vk  + \sinh \vartheta_\vk \right)  \left[i \sin (k_x a) + \sin (k_y a)\right].
\end{aligned}
\end{equation}
At corners of the Brillouin zone such as $( k_x a = \pi, \; k_y a =0 )$, the off-diagonal terms vanish and the acoustic phonon and the optical phonon are exactly degenerate. Close to these corners, the acoustic phonon band might contribute nonzero Berry curvatures, but it does not contribute to overall Chern numbers. The argument is as follows.

We can write the full Hamiltonian~(\ref{eq:four-band-Hk}) in a perturbed form,
\begin{equation}
\mathcal{H}'_\vk = \left( \begin{array}{cccc}
\epsilon_\vk^{m+} & 0 & M_1 & 0 \\
0 & \epsilon_\vk^{m-} & M_1^* &  0 \\
M_1^* & M_1 & \epsilon_\vk^{p-} & 0 \\
0 & 0 & 0 & \epsilon_\vk^{p+}
\end{array} \right) 
+ \left( \begin{array}{cccc}
0 & 0 & 0 & M_2^* \\
0 & 0 & 0 &  -M_2 \\
0 & 0 & 0 & 0 \\
M_2 & -M_2^* & 0 & 0
\end{array} \right) 
= S \left( \begin{array}{cccc}
E^{(0)}_1 &  &  &  \\
  & E^{(0)}_2 &  &  \\
  &  & E^{(0)}_3 &  \\
 &  &  & \epsilon_\vk^{p+}
\end{array} \right) S^\dagger + V'_\vk,
\end{equation}
where exact diagonalization of the upper $3 \times 3$ block yields eigenvalues $\left( E^{(0)}_1, \, E^{(0)}_2,\, E^{(0)}_3 \right)$ in an ascending order and $S = (\phi^{(0)}_1, \phi^{(0)}_2, \phi^{(0)}_3, \phi^{(0)}_4)$, where $\psi^{(0)}_n$ are column-like orthonormal eigenvectors, $\phi^{(0)}_4 = (0,0,0,1)^T$.
For any momentum away from the Brillouin zone boundary, we perform the non-degenerate perturbation theory to the first order. The three lower-energy wave functions are corrected by the presence of the fourth:
\begin{equation}
\phi^{(1)}_{n=1,2,3} = \phi^{(0)}_n + | \phi^{(0)}_4 \rangle 
\frac{\langle \phi_4 | \tilde{V}_\vk | \phi^{(0)}_n \rangle}{E^{(0)}_n - \epsilon_\vk^{p+}}.
\end{equation}
The Berry curvature is defined by
$\Omega_n^z = \partial_{k_x} A_n^y - \partial_{k_y} A_n^x$, where 
$\mathbf{A}_n = i \langle \phi_n | \nabla_\vk | \phi_n \rangle$ for a normalized wave fuction $|\phi \rangle$. For a wave function $| \tilde{\phi}_n \rangle$ that is not normalized, the Berry curvature takes the expression
\begin{equation}
\Omega_n^z
= 
i 
\frac{\langle \partial_{k_x} \tilde{\phi}_n | \partial_{k_y} \tilde{\phi}_n \rangle 
- \langle \partial_{k_y} \tilde{\phi}_n | \partial_{k_x} \tilde{\phi}_n \rangle}
{ \langle \tilde{\phi}_n | \tilde{\phi}_n \rangle} 
+ i \frac
{\langle \tilde{\phi}_n | \partial_{k_x} \tilde{\phi}_n \rangle 
\langle \partial_{k_y} \tilde{\phi}_n | \tilde{\phi}_n \rangle 
-\langle \tilde{\phi}_n | \partial_{k_y} \tilde{\phi}_n \rangle 
\langle \partial_{k_x} \tilde{\phi}_n | \tilde{\phi}_n \rangle}
{ (\langle \tilde{\phi}_n | \tilde{\phi}_n \rangle)^2} .
\end{equation}
Note $\nabla_\vk | \psi^{(0)}_4 \rangle = 0$.
Corrections to Berry curvatures from the perturbation are
\begin{equation}
\Omega^{(1)}_{n=1,2,3} = \Omega^{(0)}_{n} + \mathcal{O} 
\left(\frac{\langle \psi_4 | \tilde{V}_\vk | \psi^{(0)}_n \rangle}{E^{(0)}_n - \hbar \omega_\vk^+} \right)^2,
\end{equation}
which can be arbitrarily small as $\kappa \rightarrow 0$. Thus, there is no correction to the topological Chern number
$C_n = \int d k_x d k_y \Omega_n /2 \pi \in \mathbb{Z}$.

\section{Derivation of the continuum theory}

We now show the detailed derivation of the continuum theory.
Let $ \vn (\mathbf{r},t) \!=\! \left[\mathbf{S}_A(\mathbf{r},t) - \mathbf{S}_B(\mathbf{r},t))\right]/2S $ and
$\vm (\mathbf{r},t) = \left[\mathbf{S}_A(\mathbf{r},t) + \mathbf{S}_B(\mathbf{r},t))\right]/S$, where $\mathbf{S}_{A}(\mathbf{r},t)$~($\mathbf{S}_B(\mathbf{r},t)$) is the vector field representing spins on sublattice A~(B). The magnetic Lagrangian density is
\begin{equation}
\begin{split}
\mathcal{L}_\text{m} [\vm,\vn] &= \frac{S \hbar}{2 a^2} \dot{\vn} \cdot (\vn \times \vm) 
+ \frac{S}{2 a^2}  \mathbf{h} \cdot \vm
- \frac{J S^2}{a^2} \vm^2
- \frac{K S^2}{8 a^2} (m_x^2 + m_y^2) \\
&- (\frac{J S^2}{2} + \frac{K S^2}{8})\left[(\nabla n_x)^2 + (\nabla n_y)^2\right] 
- \frac{J S^2}{2}  (\nabla n_z)^2
- \frac{K S^2}{2 a^2} (n_x^2 + n_y^2)
\end{split}
\end{equation}
We integrate out the hard mode $\vm$, set $\mathbf{h} = h \hat{\mathbf{z}}$, and consider small deviations of $\vn(\mathbf{r},t) = \vn_0 + \delta \vn(\mathbf{r},t)$ from the N{\'e}el ground state
 $\vn_0 = \hat{\mathbf{z}}$.
Introducing complex fields $\psi^\pm = \delta n_x \pm i \delta n_y$,
\begin{equation}
\mathcal{L}_\text{m} [\psi] =
\frac{\hbar^2}{2(8J+K) a^2} \dot{\psi}^+ \dot{\psi}^-
- \frac{ i \hbar h }{16 J a^2} (\dot{\psi}^+ \psi^- - \dot{\psi}^- \psi^+)
-(\frac{J }{2} + \frac{K}{8}) S^2 (\nabla \psi^+) (\nabla \psi^-) 
-\frac{K S^2}{2 a^2} \psi^+ \psi^-.
\end{equation}
The elastic Lagrangian density is
\begin{equation}
\mathcal{L}_\text{e} = \frac{M}{2a^2} (\dot{u}^z)^2 - \frac{\lambda}{2} (\nabla u^z)^2,
\end{equation}
and the energy density of magnetoelastic coupling is
\begin{equation}
\mathcal{H}_\text{me} = - \frac{2 \kappa S^2}{a} n_0^z \vn \cdot \nabla u^z 
= - \frac{ \kappa S^2}{a}  \left[ (\psi^+ + \psi^-) \partial_x u^z - i (\psi^+ - \psi^-) \partial_y u^z \right].
\end{equation}

We perform the Fourier transform of the continuous fields as follows,
\begin{equation}
\begin{split}
\psi^+ (\mathbf{r}, t) 
&= -a \sqrt{\frac{8J+K}{\hbar \omega_m}} 
\int \frac{d^2 \vk}{2 \pi} \int \frac{d \omega}{\sqrt{2\pi}}\,
\psi^+_{\vk, \omega} e^{ i \omega t - i \vk \cdot \mathbf{r} }, \\
\psi^- (\mathbf{r}, t) &= a \sqrt{\frac{8J+K}{\hbar \omega_m}} 
\int \frac{d^2 \vk}{2 \pi} \int \frac{d \omega}{\sqrt{2\pi}}\,
\psi^-_{\vk, \omega} e^{ i \omega t - i \vk \cdot \mathbf{r} } \\
u^z (\mathbf{r}, t) &=a  \sqrt{\frac{\hbar}{2M\omega_p}} \int \frac{d^2 \vk}{2 \pi} \int \frac{d \omega}{\sqrt{2\pi}} \,
u^z_{\vk, \omega} e^{i \omega t - i \vk \cdot \mathbf{r} }.
\end{split}
\end{equation}
In the long-wavelength limit $J \gg K, h $ and $k_x^2, k_y^2 \ll K/J$,
$\hbar \omega_m \approx  S \sqrt{8JK} + S \sqrt{2J^3/K} a^2 (k_x^2 + k_y^2)$ 
and $\hbar\omega_p = \hbar\omega_0 a \sqrt{ k_x^2 + k_y^2 } $ are the dispersion relations solved from the equations of motion of $\mathcal{L}_\text{m}$ and $\mathcal{L}_\text{e}$ respectively.

Using relations $(\psi^+_{\vk,\omega})^\star = \psi^-_{-\vk,-\omega}$ 
and $u^\star_{\vk,\omega} = u_{-\vk,-\omega}$, the total action is in the form
\begin{equation}
    \mathcal{S} [\Psi] = \frac{1}{2} \int d^2 \mathbf{r} \int dt \left( \mathcal{L}_\text{m} + \mathcal{L}_\text{e} - \mathcal{H}_\text{me} \right)
    =  \frac{1}{2}\int {d^2 \vk} \int {d \omega}  \;
    \Psi_{\vk, \omega}^\dagger 
    G^{-1}_{\vk,\omega}
    \Psi_{\vk, \omega},
\end{equation}
where $\Psi_{\vk,\omega} = \left(\psi^+_{\vk,\omega}, \psi^-_{\vk,\omega}, u^z_{\vk,\omega} \right)^T$ has the dimension of [Length $\cdot$ Time], and
\begin{equation}
G^{-1}(\vk,\omega) 
= \left( \begin{array}{ccc}
     (\hbar \omega - h + \hbar \omega_m)(\hbar \omega - h - \hbar\omega_m) /2 \hbar \omega_m &
    0 &
    -\widetilde{C}_\vk \\
    0 &
    (\hbar \omega + h + \hbar \omega_m)(\hbar \omega + h - \hbar\omega_m) /2 \hbar \omega_m &
    -\widetilde{C}_\vk^* \\
    -\widetilde{C}_\vk^* &
    -\widetilde{C}_\vk &
     \hbar(\omega + \omega_p)(\omega - \omega_p)/2\omega_p
\end{array} \right),
\end{equation}
where
\begin{equation}
\widetilde{C}_\vk
\approx -\kappa S^{3/2} \sqrt{\frac{a \hbar}{2 M \omega_0}} (k_x^2 + k_y^2)^{-1/4}
(8J/K)^{1/4} (i k_x - k_y).
\end{equation}
We look for poles with positive values of the propagator $G(\vk, \omega)$ in the momentum regime where all bands are nearly degenerate 
$\hbar \omega \sim \hbar \omega_m + h \sim \hbar \omega_m - h \sim \hbar \omega_p$,
\begin{equation}
G^{-1}(\vk,\omega)
 \approx \left( \begin{array}{ccc}
    \hbar \omega - (\hbar\omega_m + h) &
    0 &
    -\widetilde{C}_\vk \\
    0 &
    \hbar \omega  - (\hbar\omega_m - h) &
    -\widetilde{C}_\vk^* \\
    -\widetilde{C}_\vk^* &
    -\widetilde{C}_\vk &
    \hbar\omega - \hbar\omega_p
\end{array} \right)
= \hbar \omega - 
 \mathcal{G}(\vk, \omega).
\end{equation}
where $\mathcal{G}(\vk, \omega)$ is defined in Eq.~(\ref{eq:propagator}) and is exactly the lowest-order approximation of the Block Hamiltonian~(\ref{eq:three-band-Hk}).
\end{document}